\documentclass{llncs}
\usepackage{graphics}
\usepackage{graphicx}
\usepackage[fleqn]{amsmath}
\usepackage{multirow}
\usepackage{subfig}
\usepackage{epsfig}
\usepackage{amssymb}
\usepackage{comment}

\begin{document}

\title{TexT - Text Extractor Tool for Handwritten Document Transcription and Annotation}

%
%
\author{Anders Hast\inst{1} \and Per Cullhed\inst{2} \and Ekta Vats\inst{1}}
\authorrunning{Anders Hast et al.} 
%
%
\institute{Department of Information Technology, Uppsala University, Uppsala, Sweden\\
\email{anders.hast@it.uu.se, per.cullhed@ub.uu.se, ekta.vats@it.uu.se}\\
\and
University Library, Uppsala University, Uppsala, Sweden}

\maketitle              

\begin{abstract}
This paper presents a framework for semi-automatic transcription of large-scale historical handwritten documents and proposes a simple user-friendly text extractor tool, $TexT$ for transcription. The proposed approach provides a quick and easy transcription of text using computer assisted interactive technique. The algorithm finds multiple occurrences of the marked text on-the-fly using a word spotting system. $TexT$ is also capable of performing on-the-fly annotation of handwritten text with automatic generation of ground truth labels, and dynamic adjustment and correction of user generated bounding box annotations with the word being perfectly encapsulated. The user can view the document and the found words in the original form or with background noise removed for easier visualization of transcription results. The effectiveness of $TexT$ is demonstrated on an archival manuscript collection from well-known publicly available dataset.
\end{abstract}


\section{Introduction}
\label{sec:intro}
When printing was invented in the mid 15th century, a sort of transcription revolution took place all over Europe. Single handwritten texts were transformed into multiple copy books. Although this invention was crucial for the growth of knowledge, the process of writing continued well into the 20th century very much as before, with the help of pen and ink.

A similar media-revolution is taking place right now when modern technology in the form of electronic texts is revolutionizing our reading habits and our media distribution possibilities. One of the most crucial steps for science in this modern media-revolution is the ability to search within texts. Optical Character Recognition (OCR) technology \cite{Mori99a,Govindan:1990:CRR:90453.90456,doi:10.1108/00220411211256021} has opened up even old printed texts to modern science in an unprecedented way. In libraries, meta-data is no longer the sole entry to collections, electronic content can speak for itself and this also changes library practices. However, the large mass of handwritten texts in our libraries and archives is still waiting to be transformed into searchable texts. The reason for this is a combination of technical and economic factors. Modern technology does not yet give us the good results of OCR technology, which nowadays can be so successfully applied to printed texts that it is a straightforward part of digitization processes world-wide. 

Handwritten text recognition (HTR) \cite{plamondon2000online,Marti:2001:USL:505741.505745,Toselli:2015:HTR:2809544.2809551,Espana-Boquera:2011:IOH:1957383.1957622,Parvez:2013:OAH:2431211.2431222} 
is an emerging field and can be quite successful in certain circumstances, especially when applied to an even and uniform handwriting, but rarely so for the non-homogeneous handwritten texts that fill our archives. In most cases, manual transcription is still the most common way to produce reliable electronic texts from handwritten texts, but modern technology advances and many projects try to tackle this problem. Manual transcription is typically expensive and prone to human error. The incentives to open up this material to computerized searches is high. The information in archives and library collections world-wide, represent an enormously important source to history and only relatively small parts of it is available as electronic texts.

Semi-automatic transcription of manuscripts typically requires hundreds of already transcribed pages, with thousands of examples of each word, in order to produce a useful transcription of the rest of the text. Due to the time consuming machine learning procedures involved, this is computed as off-line batch jobs overnight \cite{Espana-Boquera:2011:IOH:1957383.1957622}. However, this means that if just a dozen pages exist, the transcriber is forced to complete the transcriptions without the help of HTR techniques, unless a similar handwriting style exists. An alternative approach to fast transcription of text with a low cost is using computer assisted interactive techniques.

This paper introduces a simple yet effective text extractor tool, $TexT$ for transcription of historical handwritten documents. $TexT$ is designed for quick document transcription with the help of user interaction where the system finds multiple occurrences of the marked text on-the-fly using a word spotting system. Other advantages of $TexT$ include on-the-fly annotation of handwritten text with automatic generation of ground truth labels, adjustment and correction of user labeled bounding box annotations such that the word perfectly fits inside the rectangle. Nevertheless, the transcribed words are cleaned using filtering methods for background noise removal.

This paper is organized as follows. Sections \ref{sec:lit}-\ref{sec:lit3} discusses various transcription and annotation methods and tools available in literature, and discusses related work on handwritten text transcription. Section \ref{sec:method} explains the proposed text extractor tool $TexT$ in detail. Section \ref{sec:exp} demonstrate the efficacy of the proposed method with implementation details on well-known historical document dataset. Section \ref{sec:concl} concludes the paper.

\section{Transcription methods and tools}
\label{sec:lit}
Transcriptions can be made by several different techniques, by reading and typing, typically done by one person interested in using the contents of the documents, as opposed to collective transcription where many individuals make transcriptions using crowdsourcing techniques. HTR, and dictation, are other techniques that can be used to produce transcriptions. An example of the latter is the war-diary of Sven Blom, a Swedish volunteer in The Foreign Legion during the Great War. The diary is kept in Uppsala University Library and was transcribed by dictation \cite{kblink}.

Due to the labour-intensive task involved in transcriptions, crowdsourcing, a term originally coined by Jeff Howe in Wired Magazine in 2006 \cite{howe2006rise}, has been a useful way of distributing transcription work to many people and therefore it sits at the core of many successful transcription projects. The $Transcribe$ $Bentham$ project at the University College of London is often mentioned as an example \cite{moyle2011manuscript}. Like so many others, $Transcribe$ $Bentham$ is built with components from the open-source software $MediaWiki$, also used for the perhaps biggest crowdsourcing project on the planet, $Wikipedia$. $Transcribe$ $Bentham$ started in 2010 and has to this date completed approximately 43\% of the whole collection \cite{link2}. They now collaborate with the READ project \cite{link_read} and the application Transkribus \cite{link_transkribus}, which can combine HTR with manual transcription.

There are numerous other transcription tools on the Internet. Zooniverse \cite{borne2011zooniverse}, based in Oxford, include transcription as one of their crowdsourcing tasks, among many others. The plugin Scripto \cite{link3} is one of the oldest, typically created in an environment close to the history discipline, the Roy Rosenzweig Center for History and New Media at George Mason University. It is also based on $MediaWiki$ and can be used as a plugin for $Omeka$, $Wordpress$ and $Drupal$. Veele Handen \cite{link4} is a Dutch application which offers crowdsourced transcriptions as a tool for archives and libraries wishing to open up their collections. They have recently included progress bars where followers and participants can monitor progress. 

This feature is very similar to the Smithsonian Institution and their ``Digital Volunteers" \cite{link5}. In fact, the Smithsonian Institution can be regarded as one of the pioneers in assigning tasks to volunteers. Already in 1849, soon after the founding of Smithsonian Institution, it's first secretary, Joseph Henry, was able to initiate a network of some 150 volunteers for weather observations, all over the United States \cite{link6}. The ``Smithsonian Digital Volunteers" is a very successful transcription application and their Graphics User Interface (GUI) combines a clear topical structure with progress bars and a general layout which has incorporated well-established practices used in proof-reading. The work of volunteer number one, has to be approved by a second volunteer and finally the result needs to be approved by the mother institution, wishing to publish the results on the web. Together with other activities, such as promoting projects via social networks, they have managed to achieve good results, demonstrating the importance of an attractive GUI in crowdsourcing. The topical structure facilitates for the user to find attractive tasks

Uppsala university library is Sweden's oldest university library and its manu-script collections consist of approximately four kilometers of handwritten material in letters, diaries, notebooks etc. The handwritten manuscript collections date back 2000 years; from BC till the 21st century. The medieval manuscripts are plentiful and the 16th to 20th centuries are well represented with many single important collections, such as the correspondence of the Swedish King Gustav III, containing letters from, for example the French Queen Marie Antoinette and the Waller collection of 38000 manuscripts with letters from both Isaac Newton and Charles Darwin. The languages in the collection are also diverse (e.g. Swedish, Arabic, Persian etc.). However, the main languages for this project include Swedish, Latin, German, and French. 

Since a few years back it has been possible to publish digitized material in the Alvin platform \cite{link7}, a repository for cultural heritage materials shared among the universities in Uppsala, Lund and G\"oteborg, as well as other Swedish libraries and museums. However, as so often is the case, very little of the handwritten material is transcribed. The collection can therefore be accessed only through meta-data and cannot be analyzed by computational means, a problem which may only be tackled by long term and multifaceted strategic planning for producing more handwritten document transcriptions.

As a start, Alvin \cite{link7} has been adapted to allow for publishing transcriptions alongside the original manuscripts. One example of this is a transcription made from a testimony of refugees arriving to Sweden in 1945 from the concentration camp in Ravenbr\"{u}ck, kept at Lund University Library \cite{link8}. In this case, the transcriptions in textual electronic format (such as PDF) are a result of manual transcription and are open to Google indexing, thus making the original manuscripts searchable on the Internet. However, this is only an example, to open up more texts for use in digital humanities, a combination of HTR technology and manual crowdsourced transcriptions is probably as far as our present technologies admit. This work takes an initiative towards transcription and annotation of huge volumes of historical handwritten documents present in our university library using HTR methods such as word spotting \cite{hast2016segmentation}.

\section{Document annotation methods and tools}
\label{sec:lit2}
Several document image ground truth annotation methods \cite{heroux2007automatic,pletschacher2010page} and tools \cite{yanikoglu1998pink,kanungo2001trueviz,yacoub2005perfectdoc,saund2009pixlabeler,doermann2010gedi,clausner2011aletheia,biller2013webgt,valsecchi2016text} have been suggested in literature. Problems related to ground truth design, representation and creation are discussed in \cite{antonacopoulos2006ground}. However, these methods are not suitable for annotating degraded historical datasets with complex layouts \cite{2017arXiv170901775V}. For example, Pink Panther \cite{yanikoglu1998pink}, TrueViz \cite{kanungo2001trueviz}, PerfectDoc \cite{yacoub2005perfectdoc} and PixLabeler \cite{saund2009pixlabeler} work well on simple documents only and perform poorly on historical handwritten document images \cite{wei2017use}.

 A highly configurable document annotation tool GEDI \cite{doermann2010gedi} supports multiple functionalities such as merging, splitting and ordering. Aletheia \cite{clausner2011aletheia} is an advanced tool for accurate and cost effective ground truth generation of large collection of document images. WebGT \cite{biller2013webgt} provides several semi-automatic tools for annotating degraded documents and has gained importance recently. Text Encoder and Annotator (TEA) was proposed in \cite{valsecchi2016text} for manuscripts annotation using semantic web technologies. However, these tools require specific system requirements for configuration and installation. Most of these tools and methods are either not suitable for annotating historical handwritten datasets, or represent ground truths with imprecise and inaccurate bounding boxes \cite{wei2017use}. 

Our previous work \cite{2017arXiv170901775V} takes into account such issues, and proposed a simple method for annotating historical handwritten text on-the-fly. This work employs this annotation method with improvements using word spotting algorithm. A detailed discussion of the annotation tools and methods is out of scope of this paper, and the reader is referred to \cite{2017arXiv170901775V} for a deeper understanding of ground truth annotation methods, and on-the-fly handwritten text annotation in general.

\section{Related Work on Handwritten Text Transcription}
\label{sec:lit3}
Manual transcription of historical handwritten documents requires highly skilled experts, and is typically a time consuming process. Manual transcription is clearly not a feasible solution due to large amounts of data waiting to be transcribed. Fully automatic transcription using HTR techniques offers a cost-effective alternative, but often fails in delivering the required level of transcription accuracy \cite{DBLP:conf/ibpria/RomeroBHVS17}. Instead, semi-automatic or semi-supervised transcription methods have gained importance in the recent past \cite{DBLP:conf/ibpria/RomeroBHVS17,OriolRamosTerrades2010,Serrano:2009:APS:1647314.1647376,Serrano:2010:ALS:1891903.1891962,Romero:1488839}. 

The transcription method proposed in \cite{Romero:1488839} uses a computer assisted and interactive HTR technique: CATTI (Computer Assisted Transcription of Text Images) for fast, accurate and low cost transcription. For an input text line image to be transcribed, an iterative interactive process is initiated between the CATTI system and the end-user. The system thus generates successively improved transcription in response to the simple user corrective feedback. 

Image and language models from partially supervised data have been adapted in \cite{Serrano:2009:APS:1647314.1647376} to perform computer assisted handwritten text transcription using HMM-based text image modeling and n-gram language modeling. This method has been recently implemented in GIDOC (Gimp-based Interactive transcription of old text Documents) \cite{link_gidoc} system prototype where confidence measures are estimated using word graphs that helps users in finding transcription errors. 

An active learning based handwritten text transcription method is proposed in \cite{Serrano:2010:ALS:1891903.1891962} that performs a sequential line-by-line transcription of the document, and a continuously re-trained system interacts with the end-user to efficiently transcribe each line.

The performance of CATTI system \cite{Romero:1488839}, and the methods proposed in \cite{Serrano:2009:APS:1647314.1647376} and \cite{Serrano:2010:ALS:1891903.1891962} is dependent upon accurate detection of the text lines in each document page. However, the line detection and extraction in historical handwritten document images is a challenging task, and advanced line detection techniques \cite{bosch2014semiautomatic} are required. 

In practical scenarios, such methods are not appropriate as a system should ideally accept a full document page as an input and generate full transcription of the words as an output. An end-to-end system for handwritten text transcription is presented in \cite{DBLP:conf/ibpria/RomeroBHVS17,OriolRamosTerrades2010} that also uses HMM-based text image modeling with interactive computer assisted transcription. The transcription method proposed in this work addresses these issues and introduces $TexT$ for quick transcription of handwritten text using a segmentation-free word spotting algorithm \cite{hast2016segmentation}. The following section explains the proposed method and its advantages in detail. 

\section{TexT - Text Extractor Tool}
\label{sec:method}
This paper presents a framework for semi-automatic transcription of historical handwritten manuscripts and introduces a simple interactive text extractor tool, $TexT$ for transcribing words in textual electronic format. The method is based on the idea of transcribing each unique word only once for the whole document, including annotations such as gender, geographical locations, etc. This will both speed up the tedious work of transcription and also make it less exhausting. Furthermore, an interactive approach is proposed where the system finds other occurrences of the same word on-the-fly using so-called word spotting system \cite{giotis2017survey,hast2016segmentation}. The user simply identifies one occurrence, and while the word is being written by the user, the HTR engine finds other possible occurrences of the same word, which are shown to the user, meanwhile it continues in the background to search other pages. Further, the user helps the HTR engine in marking words that are correctly identified and correcting misclassified words. By marking these words, writing their corresponding letter sequence, and adding annotations, the HTR engine in the meanwhile processes these words and more accurately identifies them, making a better distinction between these two classes of words. 

The proposed method inherits features from our previous work \cite{2017arXiv170901775V} and efficiently performs on-the-fly annotation of handwritten text with automatic generation of ground truth labels, and dynamic adjustment and correction of user annotated bounding box labels with perfect encapsulation of the text inside the rectangle. Interestingly, the transcriptions are generated such that the transcribed word contains no added noise from the background or surroundings. This is made possible by the use of two band-pass filtering approach for background noise removal \cite{2017arXiv170901782V}. This is followed by connected components extraction from the word image.

The following features are important parts of the $TexT$ project planning: 
\begin{itemize}
  \item A simple yet informative, and user-friendly GUI that may attract users according to well defined topics such as botany, history, theology, diaries, etc.
  \item A GUI where the user can download the transcription results on-the-fly as they are distributed in the University library digital repository.
  \item Presence on social networks.
  \item A ranking system combined with a merit-report for the use of the contributor.
  \item A proof-reading structure with a first and a second proof-reader and a safe yet quick ingestion mechanism for the repository.
  \item A graphic illustration of progress for each topic.
  \item An administration of the application which includes active outreach to find interested audiences, close monitoring of the uploaded content and general advertising of opportunities, news and activities, including events which might give contributors extra value, such as exhibitions and shows of the original material.
  \item An HTR application, active only in the background, making use of the user input through machine learning and delivering better results based on the user input.
\end{itemize}

The combination of crowdsourcing and HTR is crucial and, it is believed to be one of the key factors for the $TexT$ project. Human interaction with AI (artificial intelligence) might be the best way to combine IT-technologies with those interested in contributing to the cultural heritage \cite{kittur2013future}. 

\begin{figure}[!ht]
\centering
\subfloat[Input document with user marked word (red) and system corrected bounding box (green).]{\includegraphics[width=0.4\linewidth]{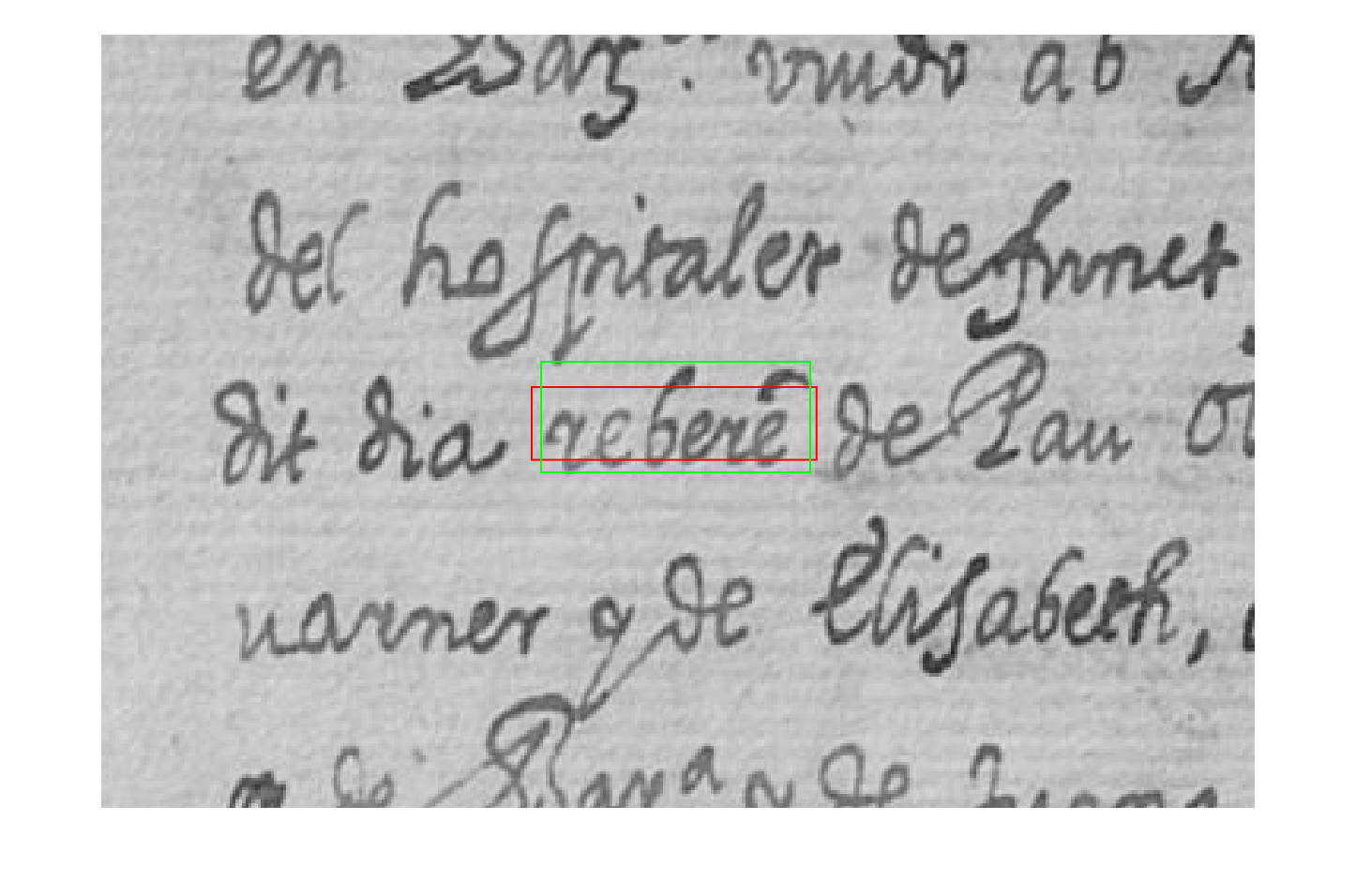}%
\label{fig_marked}} \hfil
\subfloat[Clean transcribed word {\it{reber\'e}} with background noise removed.]
{\includegraphics[width=0.4\linewidth]{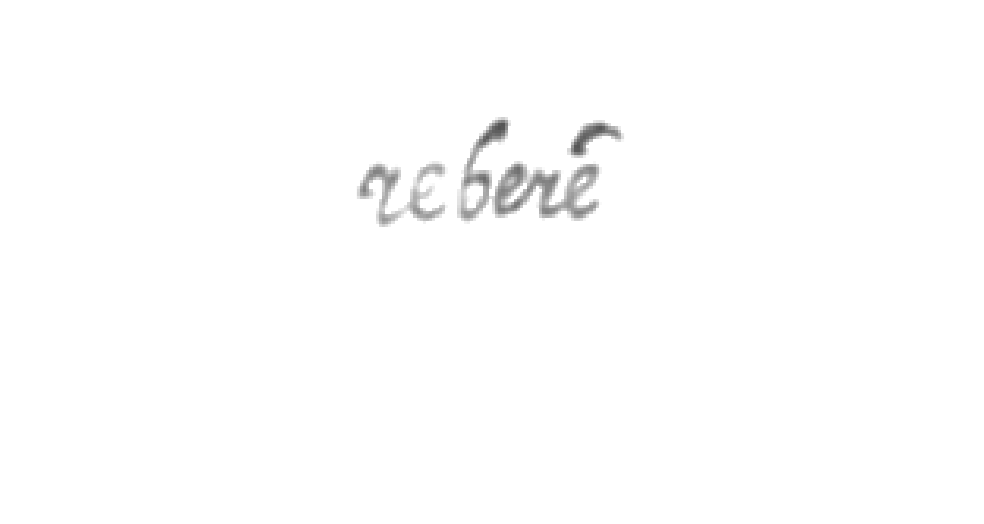}%
\label{fig_cleaned}}
\caption{The user marks a word in the document (in the left), shown in red bounding box. The system finds the best fitting rectangle (in green) to perfectly encapsulate the word. The background noise is removed and the clean transcribed word generated is shown on the right. Figure best viewed in color.}
\label{fig_mark}
\end{figure}

\section{Experimental Framework and Implementation Details}
\label{sec:exp}
This section emphasize on the overall experimental framework of $TexT$ along with insight on its implementation details. The proposed framework is tested on the Esposalles dataset \cite{romero2013esposalles}, a subset of the Barcelona Historical Handwritten Marriages (BH2M) database \cite{fernandez2014bh2m}. BH2M consists of 244 books with information on 550,000 marriages registered between 15th and 19th century. The Esposalles dataset consists of historical handwritten marriages records stored in archives of Barcelona cathedral, written between 1617 and 1619 by a single writer in old Catalan. In total, there are 174 pages handwritten by a single author corresponding to volume 69, out of which 50 pages are selected from 17th century. In future, the ancient manuscripts from the Uppsala University library will be used for further experimentation. 

\begin{figure}[!ht]
\centering
\includegraphics[width=4.5in]{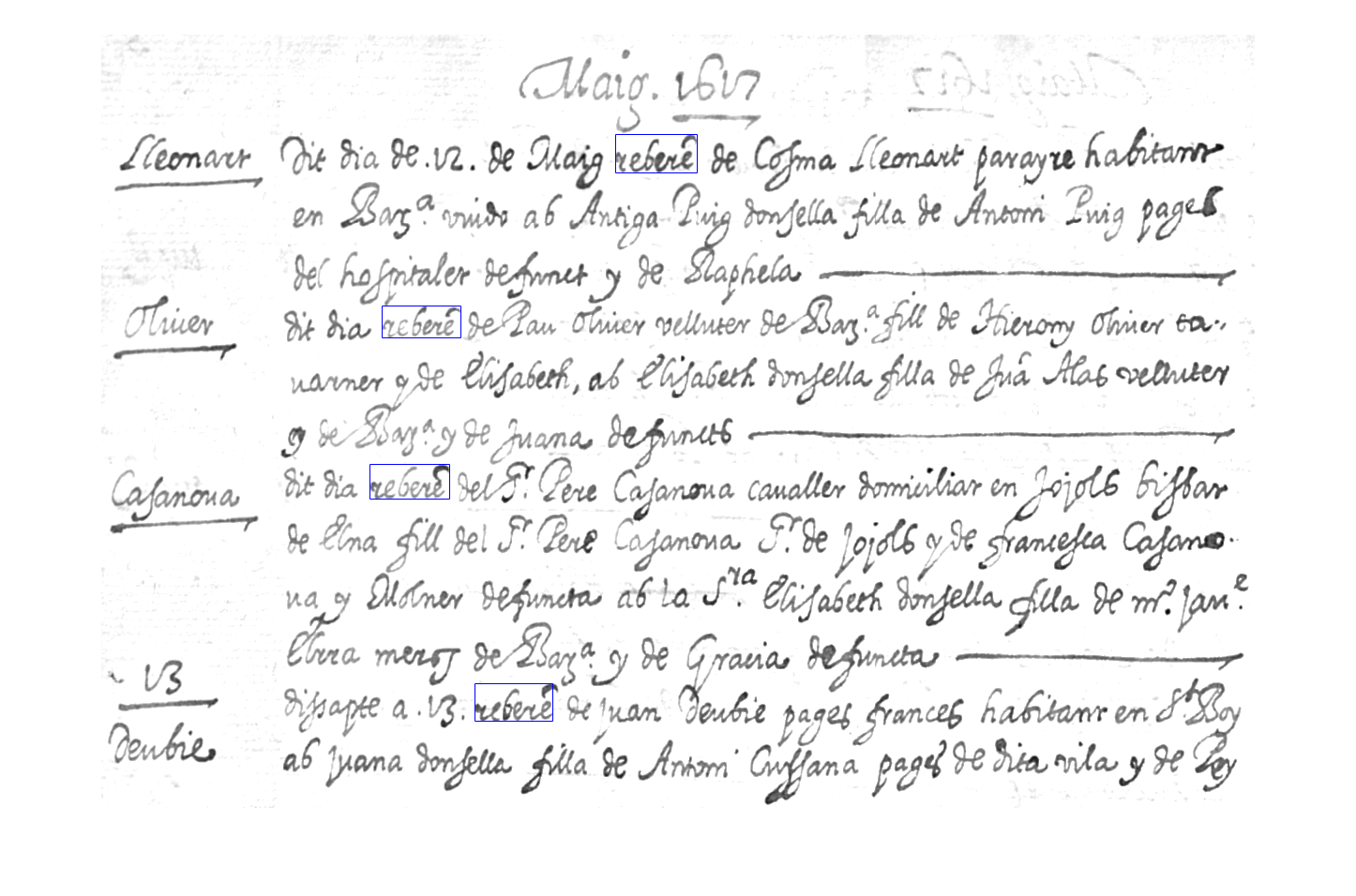}
\caption{The result of searching one word marked by the user (for example, {\it{reber\'e}}), represented using blue bounding box. Figure best viewed in color.}
\label{fig_search}
\end{figure}

\begin{figure}[!ht]
\centering
\includegraphics[width=4.5in]{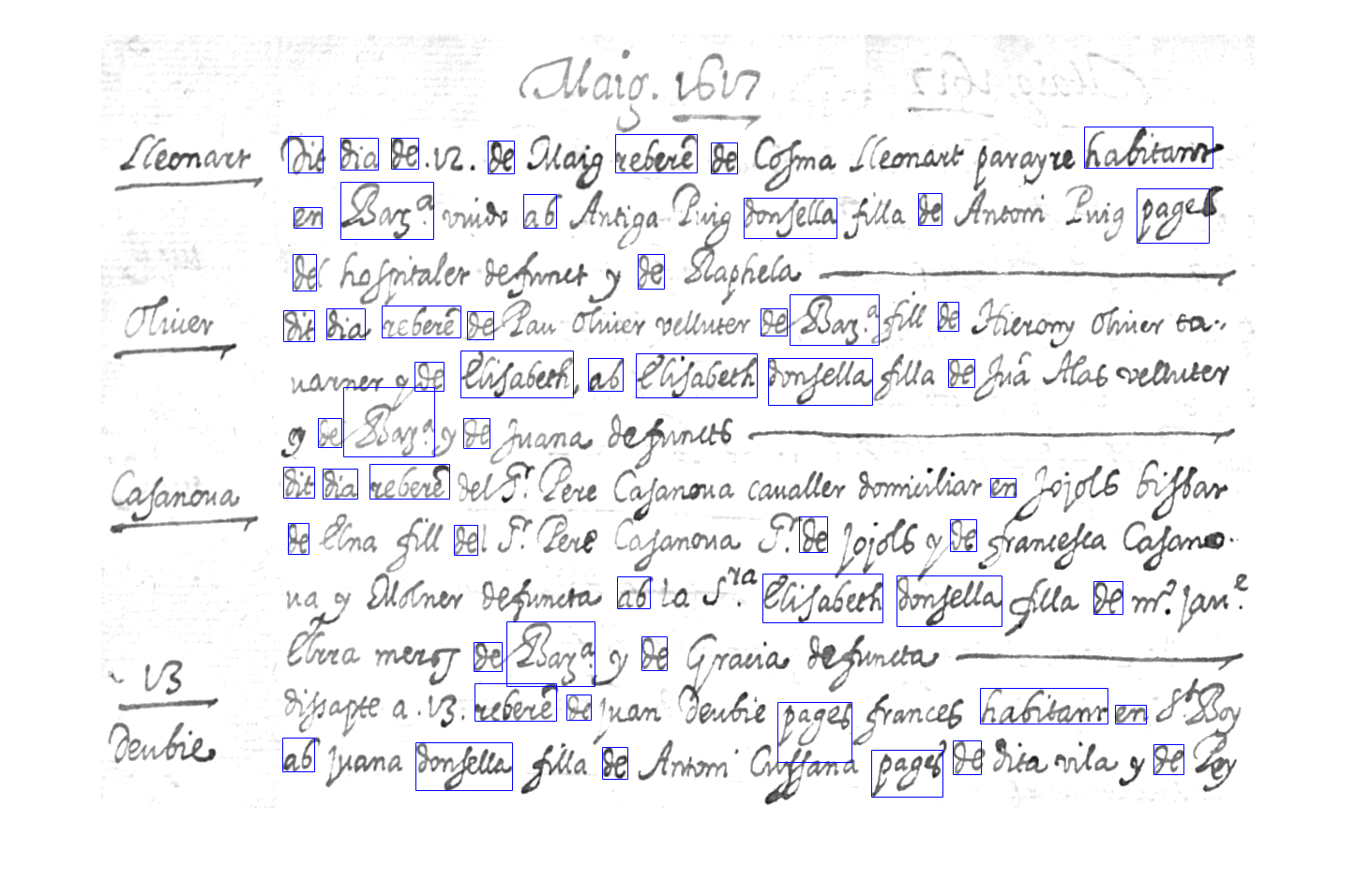}
\caption{The transcription can be performed in any order and in this case 11 different words have been marked, and the other occurrences are found automatically. Figure best viewed in color.}
\label{fig_annotated}
\end{figure}

\begin{figure}[!ht]
\centering
\includegraphics[width=4.5in]{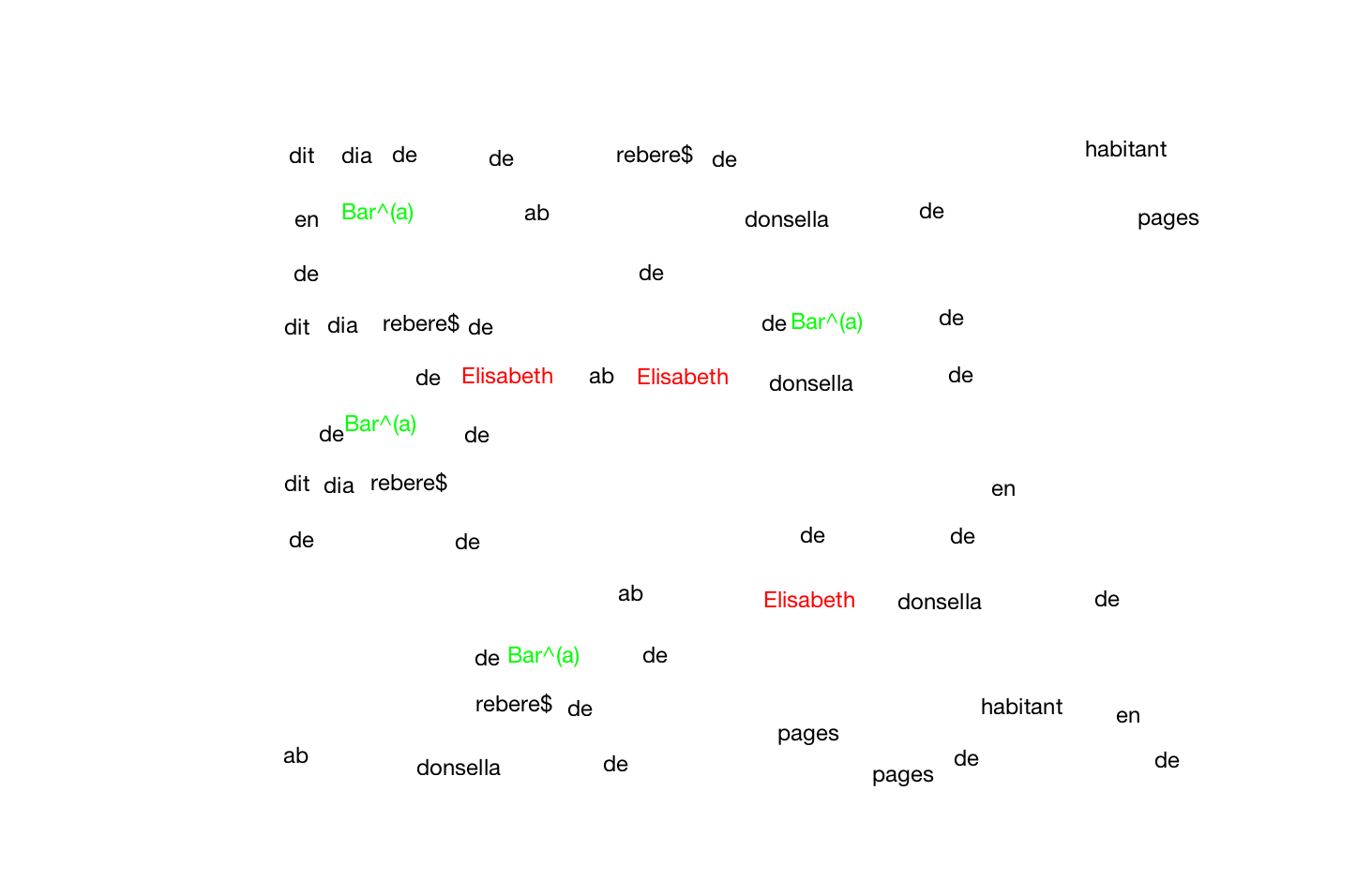}
\caption{The ongoing transcription results in words being identified in their corresponding places. In this case, the user has also annotated names and places using different colors. Figure best viewed in color.}
\label{fig:transcribed}
\end{figure}

The text transcription method based on word spotting is performed as follows. The system generates a document page query where the user marks a query word with a so called rubber band rectangle. The user marked red bounding box is highlighted in Fig. \ref{fig_marked} for a sample word {\it{reber\'e}}. The system automatically finds the best fitting rectangle to perfectly encapsulate the word, as shown in Fig. \ref{fig_marked} using green bounding box, and extracts the word. Furthermore, the noise from the background and surroundings is efficiently removed using two band-pass filtering approach in order to make the subsequent search more reliable (see Fig. \ref{fig_cleaned}). 


The system starts searching for the word in the document page and the result is shown in Fig. \ref{fig_search}. Note that only a cropped part of the document page from the dataset is shown for demonstration. The search is performed while the user inserts the transcribed text together with the annotations. Now the user can let the system learn by clicking on one or several word boxes confirming that they are correctly found. If the system find words that are misclassified, the user can inform the system by clicking a button to switch from correct to incorrect mode, and then selecting the words. While doing this, the system continues to perform word search on other document pages and update the search on the basis of information the system learns from the user.

The user can select words in any order by marking them once. Figure \ref{fig_annotated} shows how 11 words have been chosen and the system finds the rest. The corresponding transcription is shown in Fig. \ref{fig_annotated}. In this case, the user has annotated some words as names (highlighted in red) and others as geographical places (highlighted in green). This example of a place represents the abbreviation for the word $Barcelona$.

\section{Conclusion and Future Work}
\label{sec:concl}
The transcription tool $TexT$ presented in this paper is based on an interactive word spotting system, and lends itself to collaborative work, such as online crowdsourcing for large-scale document transcription. The proposed method can be further improved using client-server or cloud-based solution to perform transcription without much latency. So far algorithms for word spotting \cite{hast2016segmentation} have been developed and a simple experimental framework is proposed to support the transcription approach presented herein. 

As future work, we intend to implement a transcription framework on ancient manuscripts from Uppsala University Library that works as follows. Each user can freely mark words, annotate them and also identify words found by the search as correct or incorrect. The major part of the search will be performed on a dedicated computer that splits the work in parallel, making it possible to search even large documents in a few seconds. It can be noted that searching one word in our MATLAB implementation takes about 2 seconds for the example shown in Fig. \ref{fig_search}. The word spotting approach used in this work \cite{hast2016segmentation} efficiently performs parallel processing such that the search in a single page can be distributed into several processes, and hence making the search much faster. Different learning methods are being evaluated to improve the transcription algorithm. Deep learning techniques can be used only when several hundreds of annotated examples are available for a document, but when starting a transcription of an entirely new document, no such are usually available.

\bibliographystyle{splncs} 
\bibliography{refs}

\end{document}